# High Pressure Gases in Hollow Core Photonic Crystal Fiber: A New Nonlinear Medium

M. Azhar[1], G. K. L. Wong[1], W. Chang[1], N. Y. Joly[2,1] and P. St.J. Russell[1,2]

[1]Max Planck Institute for the Science of Light
[2]Department of Physics, University of Erlangen-Nuremberg
Günther-Scharowsky-Str. 1, 91058 Erlangen, Germany

The effective Kerr nonlinearity of hollow-core kagomé-style photonic crystal fiber (PCF) filled with argon gas increases over 100 times when the pressure is increased from 1 to 150 bar, reaching 15% of that of bulk silica glass, while the zero dispersion wavelength shifts from 300 to 900 nm. The group velocity dispersion of the system is uniquely pressure-tunable over a wide range while avoiding Raman scattering – absent in noble gases – and having an extremely high optical damage threshold. As a result, detailed and well-controlled studies of nonlinear effects can be performed, in both normal and anomalous dispersion regimes, using only a fixed-frequency pump laser. For example, the absence of Raman scattering permits clean observation, at high powers, of the interaction between a modulational instability side-band and a soliton-created dispersive wave. Excellent agreement is obtained between numerical simulations and experimental results. The system has great potential for the realisation of reconfigurable supercontinuum sources, wavelength convertors and short-pulse laser systems.

Hollow-core photonic crystal fiber (HC-PCF) offers long interaction lengths while avoiding beam diffraction, thus providing an effective environment for nonlinear optics in gases or liquids [1]. Kagomé-style HC-PCF [1] offers in addition broadband transmission with moderately low loss, features that are useful for exploring nonlinear effects (such as self-phase modulation) which generate broad-band optical spectra. It also uniquely offers a smooth pressure-tunable variation of dispersion over a very wide wavelength range, providing a perfect environment for demonstrating many different nonlinear effects, such as efficient generation of tunable deep UV light via dispersive wave generation [2] and the first observation of a plasma-related soliton blue-shift [3]. In the first of these cases the zero dispersion wavelength ($\lambda_0$) was placed closer to the pump wavelength so as to allow dispersive wave generation in the UV. In the second case it was pushed far into the UV so as to prevent dispersive-wave perturbations to soliton compression at 800 nm [3, 4]; the ensuing self-compression created peak intensities as high as $10^{14}$ W/cm$^2$, sufficient to ionize the gas.

An inherent limitation in these systems is the rather low nonlinearity provided by the gas at the pressures used. Since the efficiency of self-compression decreases significantly with soliton order $N$ [5], it is necessary to use short (~50 fs) high energy (a few µJ) pump pulses, which requires the use of a complex oscillator-amplifier laser system. In this paper, we increase the pressure in an Ar-filled kagomé PCF to 110 bar (the fibers are capable of withstanding inner pressures of at least 1000 bar), resulting in a Kerr nonlinearity that approaches that of silica glass. Uniquely, the dispersion remains low and flat from the UV to the near-IR. These features allow exploration of a range of different nonlinear regimes without changing the fiber or the pump laser. Using noble gases adds an extra twist to the system by eliminating Raman effects, allowing us to study Kerr-related phenomena in the absence of perturbations such as the soliton self-frequency shift or Raman-induced noise.

We used a kagomé HC-PCF with 18 μm core diameter, permitting $\lambda_0$ to be pressure-tuned from ~300 to ~900 nm. The experimental set-up consisted of a 28 cm length of the kagomé HC-PCF with gas cells placed at both ends. The pump laser was an amplified Ti:sapphire laser system oscillating at a center wavelength of 800 nm, delivering pulses of duration 140 fs at a repetition rate of 250 kHz. The maximum pulse energy launched into the fiber was 450 nJ. The fiber was enclosed in a steel tube joining the two gas cells, and the system could withstand pressures up to 150 bar. The diagnostics included a CCD camera, an optical spectrum analyzer and a frequency resolved optical gating (FROG) system. At a pressure of 90 bar, $\lambda_0$ coincides with the pump laser wavelength. At this pressure the nonlinear refractive index $n_2$ is only one order of magnitude lower than that of pure silica. By varying the gas pressure we are able to observe soliton fission, supercontinuum generation, dispersive wave emission and modulational instability (MI) flat relatively low pulse energies (~250 nJ).

The modal refractive index of kagomé HC-PCF is accurately approximated by that of a glass capillary [6, 7] and is given by:

$$n_{lm}(\lambda, p, T) \approx 1 + \delta(\lambda)\frac{p}{2p_0}\frac{T_0}{T} - \frac{\lambda^2 u_{lm}^2}{8\pi^2 a^2} \qquad (1)$$

where $\lambda$ is the vacuum wavelength, $\delta(\lambda)$ the Sellmeier expansion for the dielectric susceptibility of the filling gas, $p$ the gas pressure, $p_0$ atmospheric pressure, $T$ the temperature (room temperature in the experiments), $T_0 = 273$ K and $u_{lm}$ is the $m^{th}$ zero of an $(l-1)^{th}$ order Bessel function of the first kind ($l-1$ and $m$ are the azimuthal and radial orders and $a$ is the core radius). Finite element simulations and several experiments have confirmed the accuracy of this expression [5, 6-8]. Using Eq. (1), we can calculate the variation of $\lambda_0$ with gas pressure for the fundamental ($m = 1$, $l = 1$) mode of the HC-PCF used in the experiments (Fig. 1).

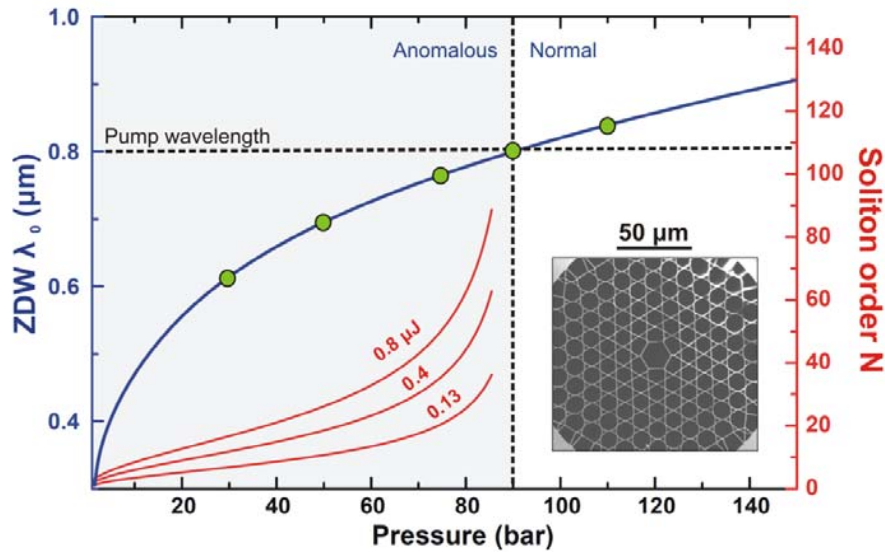

Fig. 1: Variation of $\lambda_0$ (blue trace) with Ar pressure in a kagomé HC-PCF with core diameter 18 μm. The horizontal dashed line marks the pump laser wavelength and the shaded region denotes anomalous dispersion. The green circles indicate the pressures at which experiments were performed. The red curves show how the soliton order varies with pressure for various input pump

Page | 2

energies (0.13, 0.4 ,0.8 µJ). The nonlinearity γ varies linearly from zero to $8\times10^{-5}$ $W^{-1}m^{-1}$ over the pressure range shown. The inset shows a scanning electron micrograph of the fiber structure.

It can be smoothly tuned from ~300 to 900 nm. At pressures below ~90 bar the pump wavelength lies in the anomalous dispersion regime, where soliton dynamics can be seen. The nonlinear parameter, defined by $\gamma(\lambda,p) = (d\gamma/dp)p = 2\pi(dn_2/dp)p/(\lambda A_{\text{eff}})$ where $A_{\text{eff}}$ is the effective modal area and $d\gamma/dp = 5.1\times10^{-7}$ $W^{-1}m^{-1}bar^{-1}$. At 150 bar, $\gamma = 7.65\times10^{-5}$ $W^{-1}m^{-1}$, compared to $\sim 5.2\times10^{-4}$ $W^{-1}m^{-1}$ for a silica strand with a similar core diameter.

Fig. 2 shows the experimental and theoretical output spectra with increasing input power at five different gas pressures. The expected zero dispersion wavelengths are calculated using Eq. (1) and are represented by the vertical black and white dashed lines. Remarkable agreement is found between experiment and numerical simulations using the unidirectional field propagation equation [8]. At 25 bar (Fig.2 (a)), when the pump wavelength is far from $\lambda_0$, the results are similar to those reported in [2].

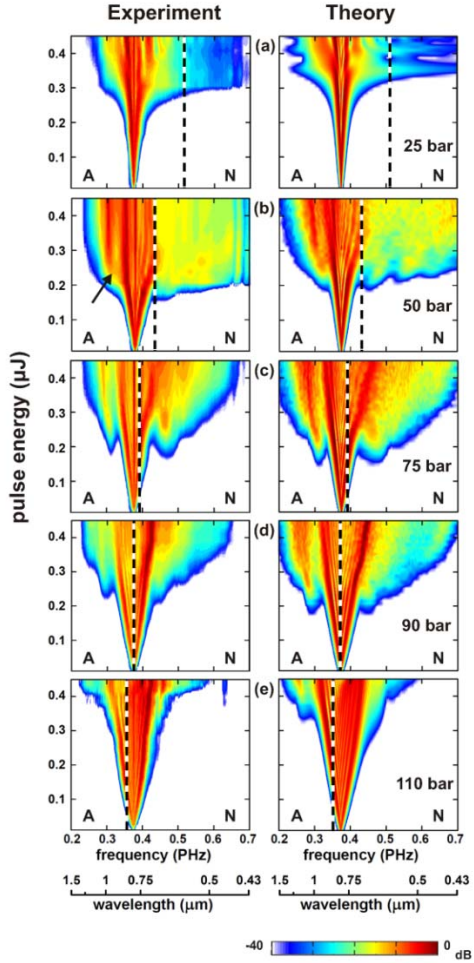

Fig. 2: Experimental and numerical evolution of the output spectra with launched pulse energy for five different gas pressures. The black and white dashed vertical lines indicate the location of $\lambda_0$. A and N denote anomalous and normal dispersion regimes. The pump frequency is kept constant at 0.375 PHz (800 nm).

The soliton order is ~27 for 450 nJ pulse energy at 25 bar ($\gamma \approx 1.34\times10^{-5}$ $W^{-1}m^{-1}$). Under these circumstances the solitons break up and phase-match to higher frequency dispersive waves in the normal dispersion regime. The emission of dispersive waves causes soliton recoil to lower frequencies, thus conserving energy [9]. Although the spectrum analyzer was unable to detect the dispersive waves directly (the simulations show that they should appear at ~300 nm), the experimental measurements show the accompanying soliton recoil at ~940 nm (~0.32 PHz). The asymmetric extension toward shorter wavelengths can be explained by the frequency-dependence of $\gamma$ [10, 11].

As the pressure is increased to 50 bar, the nonlinearity increases and $\lambda_0$ moves closer to the pump wavelength. Consequently spectral broadening appears at a much lower pump power. Numerical simulations show the appearance of multiple solitons at ~1000 nm. In the experiments too there is no evidence of any Raman-induced self-frequency shift with increasing pump power. The generated soliton remains fixed in a narrow wavelength band (indicated by the arrow on Fig. 2(b)). The spectral broadening is greater than an octave, because the low and flat dispersion profile of the gas-filled kagomé HC-PCF (compared to solid-core systems) significantly reduces group-velocity walk-off between different frequency components and ensures long interaction lengths. This enhances the effectiveness of four-wave mixing as a broadening mechanism, resulting in the generation of a cascade of sidebands, allowing it to dominate the spectral broadening process.

The very same pump laser, diagnostics and HC-PCF can be used to access a great variety of spectral broadening regimes, governed by processes such as soliton fission and MI. Such flexibility is unique to the PCF-based system.

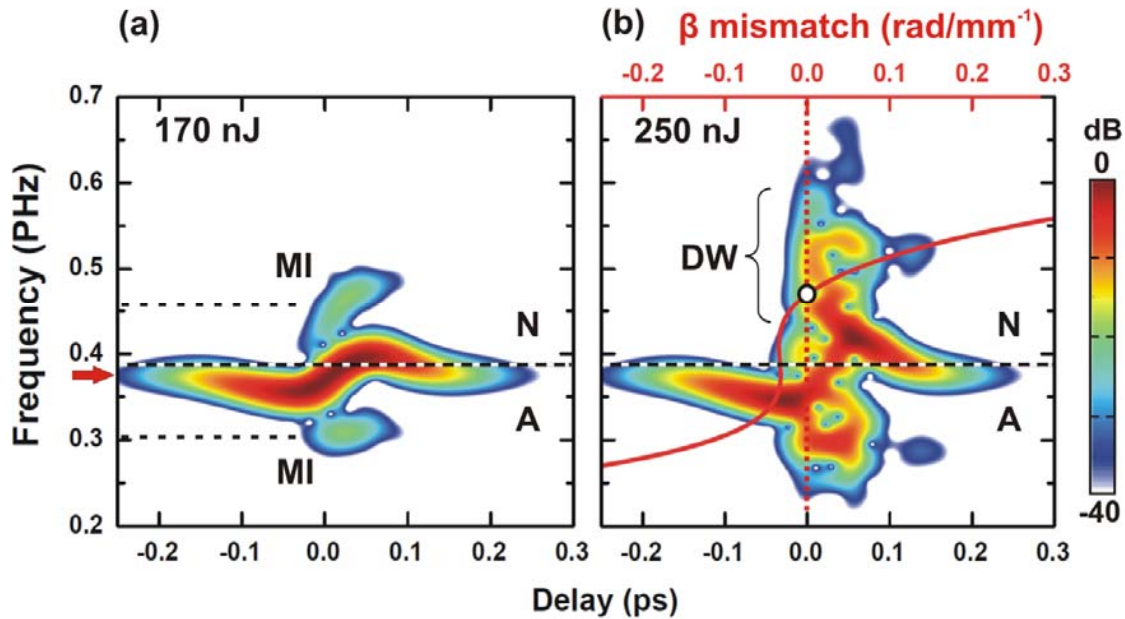

Fig. 3: Numerical X-FROG traces at 75 bar. (a) MI bands at 170 nJ launched pulse energy and (b) asymmetric spectrum at 250 nJ pulse energy when the dispersive wave band overlaps with the high frequency sideband of the MI. The black and white dashed lines indicate the position of the zero dispersion frequency $c/\lambda_0$. The red curve indicates the mismatch in propagation constant $\beta$ between the solitons and dispersive waves at a given frequency. The $\beta$ mismatch is zero at the white circle for a soliton at the pump frequency, resulting in generation of a dispersive wave in the normal dispersion regime. The red arrow marks the pump frequency.

On increasing the pressure to 75 bar (Fig. 2(c)), $\lambda_0$ moves even closer to the pump wavelength and two distinct MI sidebands can be seen at pulse energy of ~200 nJ. It is curious that these sidebands appear to be asymmetrically distant from the pump frequency. This is caused by phase-matching of the solitons to dispersive waves [9, 12], which appear only in the normal dispersion regime, creating asymmetry in the observed spectrum. A simple phase-matching analysis between the propagation constants of solitons and linear waves confirms that the dispersive wave band appears at a higher frequency than the high frequency MI side-band. This is also seen in Figs. 3(a) & (b), which show the results of a numerical X-FROG analysis of the signal after 28 cm of propagation for launched energies of 170 and 250 nJ at 75 bar. At the lower pulse energy (Figs. 3(a)), where the dispersive wave contribution is small, the MI sidebands are relatively symmetric in spacing and intensity. As the energy is increased, however, the broad dispersive wave band overlaps with the high frequency MI side-band, overwhelming it (Fig. 3(b)) and producing strong asymmetry between the side-bands. Recently Droques et al. studied the interaction between a MI side-band and a dispersive wave [12]. To avoid Raman perturbations, however, they were forced to work at low CW power levels – a limitation that is entirely absent in the PCF-based system. Of course, if required, a Raman-active gas such as hydrogen can be used if Raman effects are needed, offering yet another degree of freedom compared to existing systems.

At 90 bar and 250 nJ pulse energy, $\lambda_0$ coincides with the pump wavelength and the sidebands are symmetric in frequency (Fig. 2(d)). The X-FROG trace in Fig. 4(a) also shows a pair of distinct symmetric MI sidebands at zero delay. The self-phase modulation (SPM) trace in the X-FROG is distorted at a delay of ~110 fs, due to the effects of higher order dispersion, which become important close to $\lambda_0$. Using numerical simulations, we followed the propagation of the pulse over a longer length (58.5 cm). Both the dispersive wave band (overlapping with the high frequency MI band) and the soliton regime (anomalous dispersion) are clearly seen in Fig. 4(b). At 110 bar ($\gamma \approx 5.65 \times 10^{-5}$ W$^{-1}$m$^{-1}$, Fig 2(e)), the dispersion is normal and spectral broadening due to SPM is observed.

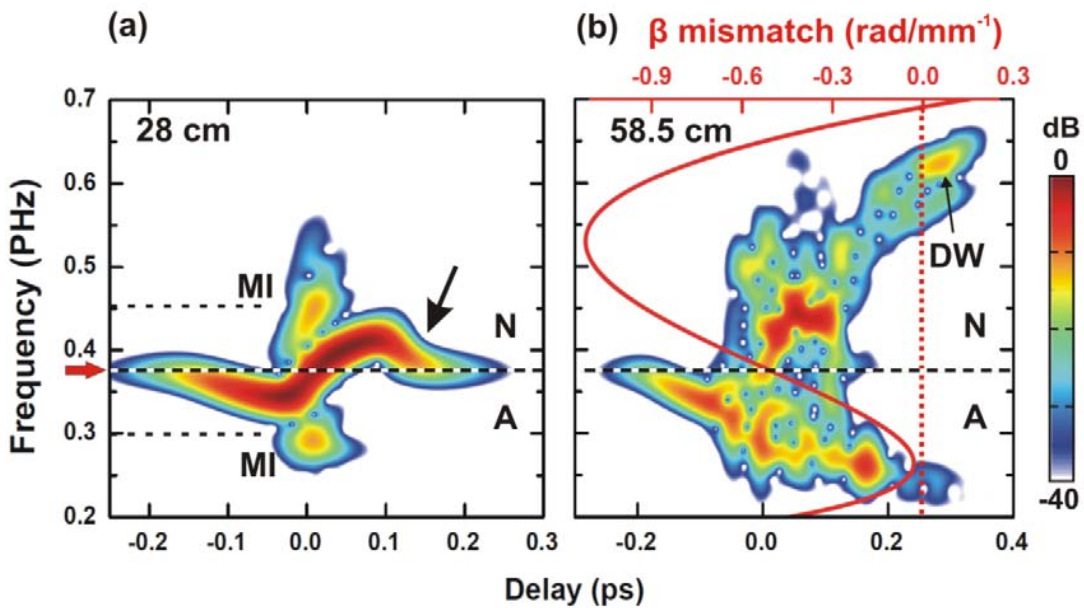

Fig. 4: Numerical X-FROG traces for 250 nJ launched pulses at 90 bar. (a) Symmetric MI spectrum after 28 cm of propagation. Arrow denotes the asymmetric SPM due to higher order dispersion. (b) after 58.5 cm. Dotted lines have same meaning than for Fig. 3.

In conclusion, noble gases at high pressure can be introduced into kagomé-style HC-PCF, providing a unique and highly versatile single-mode fiber system for exploring gas-based nonlinear optics in the absence of Raman scattering. The pressure-tunable system allows studies of nonlinear dynamics in different dispersion landscapes and over a wide range of different nonlinearity levels and soliton orders. Regimes of normal and anomalous dispersion can be readily accessed by a fixed-frequency laser merely by tuning the gas pressure. The gas-filled hollow core allows very high energies to be launched (if needed) without optical damage or photo-darkening – serious problems in fibers with solid glass cores. The system is simple and remarkable agreement can be reached between experiment and numerical simulations based on the unidirectional field propagation equation. High nonlinearity and normal dispersion at pressures above 90 bar, together with the absence of Raman scattering, make this system a promising candidate for novel studies of nonlinear optics. It could also be important for the generation of correlated photon pairs by eliminating the problem of Raman-generated noise [13]. Other noble gases may also be used. Xe, which has a nonlinearity ~30 times higher than Ar, would reach the same nonlinearity as silica at a pressure of only ~50 bar. The results pave the way for a new series of experiments on ultrafast nonlinear dynamics in highly nonlinear Raman-free systems. Hollow-core kagomé-style PCF, filled with gases at very high pressure, allows us for the first time to elevate gases to the status of "honorary solid state materials", with the added advantages of tunable dispersion and extremely high optical damage resistance.